\documentclass[reprint,amsmath,amssymb,aps,pra,superscriptaddress,longbibliography]{revtex4-2}

\usepackage{graphicx}
\usepackage{xcolor}
\usepackage{dcolumn}
\usepackage{bm}
\usepackage{physics}
\usepackage{braket}
\usepackage{hyperref}
\hypersetup{colorlinks=true, citecolor=blue, urlcolor=blue, linkcolor=blue, unicode=true}

\newcommand{\di}{\mathrm{d}}
\renewcommand{\vec}[1]{\mathbf{#1}}

\newcommand{\su}{\mathfrak{su}}
\newcommand{\SU}{\mathrm{SU}}

\newcommand{\appref}[1]{appendix~\ref{#1}}

\newcommand{\secref}[1]{section~\ref{#1}}

\newcommand{\figref}[1]{Fig.~\ref{#1}}
\newcommand{\Figref}[1]{Figure~\ref{#1}}
\renewcommand{\eqref}[1]{Eq.~(\ref{#1})}
\newcommand{\Eqref}[1]{Equation~(\ref{#1})}

\begin{document}

\title{Quantum dynamics of a fully-blockaded Rydberg atom ensemble}

\author{Dominik S.~Wild}
\affiliation{Max-Planck-Institut für Quantenoptik, 85748 Garching, Germany}
\affiliation{QuEra Computing Inc., 1284 Soldiers Field Road, Boston, MA 02135, USA}

\author{Sabina Dr\u{a}goi}
\affiliation{QuEra Computing Inc., 1284 Soldiers Field Road, Boston, MA 02135, USA}
\affiliation{Department of Physics, Harvard University, Cambridge, MA 02138, USA}

\author{Corbin McElhanney}
\thanks{Current affiliation: Snowflake Inc., Bozeman, MT, 59715, USA}
\affiliation{QuEra Computing Inc., 1284 Soldiers Field Road, Boston, MA 02135, USA}

\author{Jonathan Wurtz}
\affiliation{QuEra Computing Inc., 1284 Soldiers Field Road, Boston, MA 02135, USA}

\author{Sheng-Tao Wang}
\email{swang@quera.com}
\affiliation{QuEra Computing Inc., 1284 Soldiers Field Road, Boston, MA 02135, USA}

\begin{abstract}
    Classical simulation of quantum systems plays an important role in the study of many-body phenomena and in the benchmarking and verification of quantum technologies. Exact simulation is often limited to small systems because the dimension of the Hilbert space increases exponentially with the size of the system. For systems that possess a high degree of symmetry, however, classical simulation can reach much larger sizes. Here, we consider an ensemble of strongly interacting atoms with permutation symmetry, enabling the simulation of dynamics of hundreds of atoms at arbitrarily long evolution times. The system is realized by an ensemble of three-level atoms, where one of the levels corresponds to a highly excited Rydberg state. In the limit of all-to-all Rydberg blockade, the Hamiltonian is invariant under permutation of the atoms. Using techniques from representation theory, we construct a block-diagonal form of the Hamiltonian, where the size of the largest block increases only linearly with the system size. We apply this formalism to derive efficient pulse sequences to prepare arbitrary permutation-invariant quantum states. Moreover, we study the quantum dynamics following a quench, uncovering a parameter regime in which the system thermalizes slowly and exhibits pronounced revivals. Our results create new opportunities for the experimental and theoretical study of large interacting and nonintegrable quantum systems.
\end{abstract}

\maketitle

\section{\label{sec:sec1}Introduction}

Characterizing and understanding the emergent behavior of quantum many-body systems has been a key research focus of condensed-matter physics. Many questions still remain unanswered owing to the complexity of these systems. Recent experimental advancements have created new avenues to approach these open problems. It is now possible to experimentally study the dynamics of large quantum systems and to directly probe physical phenomena such as the growth of entanglement and the presence or absence of thermalization~\cite{schreiber2015observation, gonzalo2015localization, islam2015measuring, smith2016many, xu2018emulating, brydges2019probing, lukin2019probing, ueda2020quantum, chiaro2022direct}. The high level of precise control required for these experiments moreover constitutes a crucial step towards building large-scale quantum computers.

Ultracold atoms have been a particularly fruitful platform for this line of research~\cite{bloch2008many,bloch2012quantum,gross2017quantum,gross2021quantum,kaufman2021quantum}. By means of cooling and trapping, one can prepare large ensembles of identical atoms. The internal degrees of freedom of each atom, such as distinct hyperfine levels, give rise to a discrete state space suitable for quantum information processing. While the states can be manipulated by external electromagnetic fields, they also retain coherence for a long time due to their weak coupling to the vacuum. The interaction between atoms in low-lying states is typically weak, which is beneficial for preserving the collective quantum state, but prevents the creation of complex states with large entanglement.

\begin{figure}
    \centering
    \includegraphics{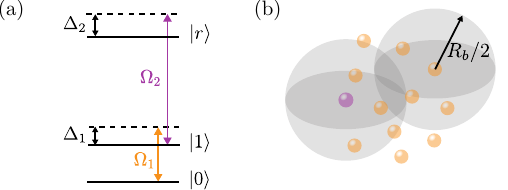}
    \caption{System setup. (a)~We consider atoms with two hyperfine ground states $\ket{0}$ and $\ket{1}$ and a highly excited Rydberg state $\ket{r}$. The hyperfine states are coupled by a driving field with Rabi frequency $\Omega_1$ and detuning $\Delta_1$. The state $\ket{1}$ is further coupled to $\ket{r}$ with Rabi frequency $\Omega_2$ and detuning $\Delta_2$. (b)~The strong van der Waals interaction between two Rydberg states defines a blockade radius $R_b$, within which the excitation of more than two atoms to the Rydberg state is effectively forbidden. We assume that $R_b$ is greater than the largest distance between any pair of atoms. Equivalently, a sphere of radius $R_b/2$ around an atom in the Rydberg state (purple) is assumed to intersect the sphere of the same size centered at any other atom (orange).}
    \label{fig:rydbergblockade}
\end{figure}

This limitation can be overcome, for example, by promoting atoms to highly-excited Rydberg states, which interact strongly via the van-der-Waals interaction~\cite{saffman2010quantum}. In this work, we investigate the quantum dynamics of the collective many-body system in the regime of the so-called \emph{Rydberg blockade}, where the van-der-Waals interaction is so strong that it prevents excitation of more than one atom within a certain volume~\cite{jaksch2000fast,lukin2001dipole,urban2009observation}. In practice, a fully-blockaded ensemble may contain hundreds of atoms~\cite{xu2021fast}. When trapped individually in an array of optical tweezers, typical values of the blockade radius and the separation between atoms allow for a fully blockaded system comprising tens of atoms~\cite{labuhn2016tunable,ebadi2021quantum,graham2022multi,ebadi2022quantum,wurtz2023aquila}. The Rydberg blockade realizes an effective multi-atom interaction, which can be used to directly implement few-qubit gates without a decomposition into two-qubit gates~\cite{isenhower2011multibit, levine2019parallel, khazali2020fast, dlaska2022quantum, jandura2022time, evered2023high}.

To advance the development of quantum technologies based on Rydberg atoms, it is desirable to develop classical simulations that enable efficient benchmarking. Owing to its nonlocality, the Rydberg blockade is expected to lead to rapid growth of entanglement. Many classical simulation techniques such as tensor networks will therefore be limited to short times or small systems. We address this challenge by considering a constrained model with a high degree of symmetry. In particular, we analyze instances where all atoms are within the blockade radius of each other, as shown by Fig.~\ref{fig:rydbergblockade}(b), and all driving fields are homogeneous. Due to the separation of scales inherent to the Rydberg blockade and thus insensitivity to the exact value of the interaction, the system is invariant under any permutation of the atoms. This symmetry enables numerical simulation of the quantum dynamics of ensembles containing several hundreds of atoms for arbitrarily long times.

Our formal approach is based on a representation theoretical analysis of the quantum Hamiltonian, which we introduce in \secref{sec:setup}. The analysis may be viewed as a generalization of Dicke states from two-level to three-level atoms~\cite{dicke1954coherence}. Dicke states correspond to eigenstates of total angular momentum, which are mathematically constructed as the irreducible representations of the Lie group $\SU(2)$. For atoms with three levels, we have to instead consider
the irreducible representations of $\SU(3)$. We apply this formalism in \secref{sec:preparation} to construct pulse sequences that prepare arbitrary permutation-invariant quantum states. In \secref{sec:equilibration}, we investigate the dynamics of a large system initialized in product states. We uncover a parameter regime in which the system exhibits robust revivals. The revivals indicate a slow progression to equilibrium despite the strong interactions. We show that the revivals are intimately linked to the symmetries of the model and provide a simple explanation in terms of an effective spin model. We summarize our findings in \secref{sec:summary} and discuss directions for future research.

\section{Setup and notation\label{sec:setup}}

\subsection{System Hamiltonian}

Throughout this work, we consider a set of $n$ identical atoms whose dynamics are restricted to three internal states: two hyperfine states denoted by $\ket{0}$ and $\ket{1}$, and a highly excited Rydberg state $\ket{r}$. Transitions between these states are controlled by two external driving fields as shown in \figref{fig:rydbergblockade}(a). The first drive (Rabi frequency $\Omega_1$, detuning $\Delta_1$) acts on the $\ket{0} \leftrightarrow \ket{1}$ transition, while the second drive ($\Omega_2$, $\Delta_2$) acts on $\ket{1} \leftrightarrow \ket{r}$. Both drives are assumed to be global, meaning that the Rabi frequencies and detunings are the same for each atom.

The large polarizability of the Rydberg state gives rise to strong van der Waals interactions between nearby atoms, which creates an energy shift if two atoms are in the Rydberg state. The energy shift scales with the sixth power of the inverse distance~\cite{saffman2010quantum}. At short distances, the energy shift can therefore be much larger than any other scale, in which case states with more than one Rydberg excitation can be adiabatically eliminated. This mechanism is the so-called Rydberg blockade. The adiabatic elimination is valid at distances shorter than the blockade radius, $R_b$, which depends on the driving strength~\cite{saffman2010quantum}. We assume that $R_b$ exceeds the largest separation between any pair of atoms (see \figref{fig:rydbergblockade}(b)), so that it is a good approximation to assume that no more than one Rydberg excitation exists in the system. We will thus restrict our analysis to the subspace containing zero or one Rydberg excitation. The Hamiltonian describing the system can then be written, in the rotating frame and within the rotating-wave approximation, as
\begin{align}
    \label{eq:Hamiltonian1}
    H &= - \Delta_1 \sum_{i=1}^n |1\rangle \langle 1|_i - (\Delta_1 + \Delta_2) \sum_{i=1}^n |r\rangle \langle r|_i \\
    &+ \frac{1}{2} \sum_{i=1}^n (\Omega_1 |1\rangle \langle 0|_i + \text{h.c.} ) + \frac{1}{2} P \sum_{i=1}^n (\Omega_2 |r\rangle \langle 1|_i + \text{h.c.} ) P, \nonumber
\end{align}
where the subscript $i$ indicates which atom an operator acts on and $P$ projects onto the subspace with at most one Rydberg excitation.

As a consequence of the global drive and the insensitivity of the Rydberg blockade to the distance, the Hamiltonian is invariant under permutations of the atoms. To take advantage of this symmetry, we introduce collective raising operators
\begin{align}
    T^+ &= \sum_{i = 1}^n | 0\rangle \langle 1 |_i, \label{eq:t+}\\
    U^+ &= \sum_{i = 1}^n | 1\rangle \langle r |_i, \label{eq:u+}\\
    V^+ &= \sum_{i = 1}^n | 0\rangle \langle r |_i \label{eq:v+}.
\end{align}
We further define the lowering operator $T^- = (T^+)^\dagger$ and the Hermitian operators $T^x = (T^+ + T^-) / 2$, $T^y = (T^+ - T^-) / (2i)$, and $T^z = \comm{T^+}{T^-}/2$ with analogous expressions for $U$ and $V$. Together, these operators may be viewed as a generalization of collective spin operators to the setting of three-level systems. The Hamiltonian can be rewritten in terms of these collective operators as 
\begin{align}
     H &= \frac{2}{3} \Delta_1 T^z + \frac{2}{3} \Delta_2 U^z + \frac{2}{3} (\Delta_1 + \Delta_2) V^z \nonumber\\
     & \quad + \frac{1}{2}(\Omega_1 T^- + \Omega_1^* T^+) + \frac{1}{2} P (\Omega_2 U^- + \Omega_2^* U^+) P,
     \label{eq:Hamiltonian2}
\end{align}
where we dropped an unimportant term proportional to the identity.

\subsection{\texorpdfstring{Review of irreducible representations of $\su(3)$}{Review of irreducible representations of su(3)} \label{sec:review}}

The operators $T^\alpha$, $U^\alpha$, and $V^\alpha$ ($\alpha \in \{ x, y, z\}$) generate the Lie algebra $\su(3)$ of the special unitary group $\SU(3)$. We can see this by considering a single atom, for which these operators are represented by traceless, Hermitian $3 \times 3$ matrices. Moreover, any traceless, Hermitian $3 \times 3$ matrix can be expressed as a linear combination of these operators with real coefficients. Hence, the algebra spanned by the operators corresponds to the defining representation of $\su(3)$. We note that the 9 operators are overcomplete since the Lie group $\SU(3)$ is only 8 dimensional. Indeed, one of the operators is redundant due to the linear relation $T^z + U^z = V^z$.

The Lie algebra generated by a set of operators is fully determined by their commutation relations. We provide the full set of commutation relations between $T^\alpha$, $U^\alpha$, and $V^\alpha$ in \appref{app:commutators}. Since the commutation relations are independent of the number of atoms, these operators also describe a representation of $\su(3)$ for more than one atom. The representation is reducible except for the special case of a single atom. Our strategy is to decompose the representation into a direct sum of irreducible representations, in which the operators $T^\alpha$, $U^\alpha$, and $V^\alpha$ become block diagonal. We will see that the projector $P$, and hence the Hamiltonian, is block diagonal in the same basis. This greatly simplifies the computation of many quantities such as the eigenenergies. Formally, the block-diagonal form exists as a consequence of Schur--Weyl duality for permutation-invariant systems (see \appref{app:schur_weyl}). In what follows, we pursue a less abstract viewpoint and focus on an explicit construction of the matrix representation of the irreducible blocks of $H$.

Before we turn to the irreducible representations of $\su(3)$, let us review the simpler case of $\su(2)$. This is particularly instructive because the operators $T^\alpha$ generate a $\su(2)$ subalgebra of $\su(3)$. This follows from the fact that for a single atom, these operators act on the two-dimensional subspace spanned by $\{ \ket{0}, \ket{1} \}$. Alternatively, one can verify that they satisfy the familiar commutation relations of angular momentum, $\comm{T^\alpha}{T^\beta} = i \sum_{\gamma \in \{x, y, z\}} \varepsilon_{\alpha \beta \gamma} T^\gamma$, where $\varepsilon_{\alpha \beta \gamma}$ is the Levi-Civita symbol. Similar arguments apply to the subalgebras generated by $U^\alpha$ and $V^\alpha$.

\begin{figure}
    \centering
    \includegraphics{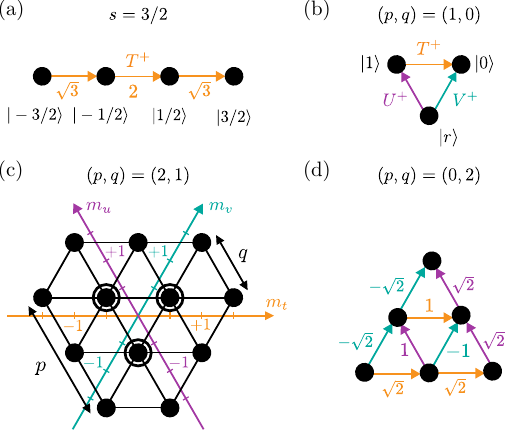}
    \caption{Irreducible representations of $\su(2)$ and $\su(3)$.
    (a)~Weight diagram of the spin $s=3/2$ representation of $\su(2)$. Each dot corresponds to a basis state $\ket{m}$. The raising operator $T^+$ induces transitions between the states as indicated by the orange arrows. The numbers next to the arrows give the associated matrix element.
    (b)~The three levels of a single atom form the $(p, q) = (1, 0)$ representation of $\su(3)$. The colored arrows show the directions in which the three distinct raising operators act.
    (c)~In general, the weight diagram of the $(p, q)$ representation of $\su(3)$ takes the form of a hexagon with side lengths $p$ and $q$. Each solid dot and each surrounding circle correspond to a basis state. The basis states are simultaneous eigenstates of $T^z$, $U^z$, and $V^z$. The respective eigenvalues $m_t$, $m_u$, and $m_v$ can be read off at the orthogonal projection onto the corresponding coordinate axis. The origin of the axis lies at the center of the diagram.
    (d)~The matrix elements of the raising operators $T^+$, $U^+$, and $V^+$ for $(p,q)=(0,2)$ and the same color coding as in (b).}
    \label{fig:weight_diagrams}
\end{figure}

The finite-dimensional, irreducible representations of $\su(2)$ are commonly labeled by the spin $s$, which is a non-negative integer multiple of $1/2$. The dimension of the representation is $2 s + 1$. There exists an orthonormal basis $\{ \ket{m} \}$, where $m$ runs from $-s$ to $s$ in integer increments. In this basis, the matrix elements of the generators of the Lie algebra, or spin operators, are fully determined by the relations $T^\pm \ket{m} = \sqrt{s(s+1) - m(m \pm 1)} \ket{m \pm 1}$ and $T^z \ket{m} = m \ket{m}$. The basis states belonging to an irreducible representation are commonly illustrated using a so-called weight diagram~\cite{greiner1989quantum,hall2015lie}, shown in \figref{fig:weight_diagrams}(a). Each dot represents a basis state $\ket{m}$, ordered from left to right by increasing $m$. Adjacent states are connected by nonvanishing matrix elements of $T^\pm$.

The above concepts set a stage for our discussion of the representations of $\su(3)$. We restrict ourselves to the aspects most pertinent to our problem and refer the reader to references~\cite{greiner1989quantum,hall2015lie} for a more comprehensive exposition. To describe the irreducible representations, one can generalize the weight diagrams introduced above. For $\su(3)$, the nodes of the weight diagrams occupy a two-dimensional triangular lattice, in contrast to the one-dimensional lattice in the case of $\su(2)$. The raising and lowering operators $T^\pm$, $U^\pm$, and $V^\pm$ induce transitions between neighboring nodes along the three distinct lattice directions. The representation corresponding to the three basis states of a single atom hence form a triangle as shown in \figref{fig:weight_diagrams}(b). 

Two further examples of weight diagrams of irreducible representations of $\su(3)$ are shown in \figref{fig:weight_diagrams}(c) and (d). In general, each irreducible representation forms a hexagon that can be labeled by two nonnegative integers $(p, q)$. Three of the sides of the hexagon have length $p$, the other three length $q$. If either $p=0$ or $q=0$, the hexagon simplifies to a triangle (e.g., \figref{fig:weight_diagrams}(b) and (d)). When $p = q = 0$, the weight diagram consists of a single node. The basis states associated with each node are indicated by black dots and circles. A single black dot means that there is one state associated with this node. Every circle surrounding a dot corresponds to an additional, degenerate basis state. The outermost nodes are always singly occupied. Moving inwards, the number of states per node increases by one if the nodes in the previous step formed a hexagon; if they form a triangle, the number of states stays the same. Following these rules, the $(p, q)$ representation can be shown to have dimension $(p+1)(q+1)(p+q+2)/2$~\cite{greiner1989quantum}.

The basis states are simultaneous eigenstates of $T^z$, $U^z$, and $V^z$. They are arranged in the weight diagram in such a way that the operators $T^\pm$, $U^\pm$, and $V^\pm$ induce transitions between states occupying adjacent nodes along the three distinct lattice directions. Consequently, we can introduce three coordinate axes that determine the eigenvalues $m_t$, $m_u$, and $m_v$ respectively belonging to $T^z$, $U^z$, and $V^z$ at each point of the weight diagram as shown in \figref{fig:weight_diagrams}(c). We can hence obtain the matrix elements of the diagonal operators $T^z$, $U^z$, and $V^z$ simply from the layout of the basis states in the weight diagram.

To compute the matrix elements of the raising and lowering operators $T^\pm$, $U^\pm$, and $V^\pm$, we make use of the fact that the states along a single line in a lattice direction form the basis of a representation of $\su(2)$. It follows that the matrix elements of the operators acting along these lines obey the rules of standard spin operators. However, care must be taken when constructing the matrix elements along different directions, especially when there is more than one state per node, to ensure that all commutation relations of $\su(3)$ are satisfied. We present a general recipe to compute the matrix elements relevant for our purposes in \appref{app:matrix_elements}. As an illustration, \figref{fig:weight_diagrams}(d) shows the matrix elements of the raising operators $T^+$, $U^+$, and $V^+$ in the $(p,q) = (0,2)$ representation. While the magnitude of the matrix elements can be understood by viewing each line along a lattice direction as either a spin-$1/2$ or spin-$1$ system, some matrix elements are inevitably negative because the operators $T^+$, $U^+$, and $V^+$ do not mutually commute.

\subsection{\texorpdfstring{Representation of $n$ atoms}{Representation of n atoms}}\label{sec:natoms}

Following the above steps, we can construct explicit matrix representations of the operators $T^\alpha$, $U^\alpha$, and $V^\alpha$ for any irreducible representation $(p, q)$. In this section, we describe how the reducible representation of $n$ atoms, defined by Eqs.~(\ref{eq:t+})--(\ref{eq:v+}), decomposes into these irreducible representations. While $p$ and $q$ can in principle be any nonnegative integers, the values that arise for $n$ atoms are restricted. This is similar to the addition of spins, where $n$ spin-$1/2$ systems can combine into a total spin with quantum number of at most $n / 2$. Given $n$ atoms with three levels, the allowed values of $p$ and $q$ are in one-to-one correspondence with partitions of $n$ into a sum of three non-negative integers~\cite{greiner1989quantum,hall2015lie}. More concretely, let $\lambda_1 \geq \lambda_2 \geq \lambda_3 \geq 0$ be three integers such that
\begin{equation}
    \lambda_1 + \lambda_2 + \lambda_3 = n.
\end{equation}
Each such partition of $n$ corresponds to an allowed irreducible representation with
\begin{equation}
    p = \lambda_1 - \lambda_2, \qquad q = \lambda_2 - \lambda_3.
\end{equation}
In what follows, we will refer to an irreducible representation interchangeably by the labels $(p, q)$ or by a partition $\lambda = (\lambda_1, \lambda_2, \lambda_3)$.

The Rydberg blockade places further constraints on the relevant irreducible representations by limiting the allowed basis states within each sector to have at most one Rydberg excitation. The number of atoms in the $\ket{0}$, $\ket{1}$, and $\ket{r}$ states is well defined for every node in a weight diagram. This follows from expressing the number operators as
\begin{align}
    \label{eq:n0}
    n_0 &= \sum_{i=1}^n \ket{0} \bra{0}_i = \frac{1}{3} \left(n \mathbb{I} + 2 T^z + 2 V^z \right),\\
    \label{eq:n1}
    n_1 &= \sum_{i=1}^n \ket{1} \bra{1}_i = \frac{1}{3} \left(n \mathbb{I} - 2 T^z + 2 U^z \right), \\
    \label{eq:nr}
    n_r &= \sum_{i=1}^n \ket{r} \bra{r}_i = \frac{1}{3} \left(n \mathbb{I} - 2 U^z - 2 V^z \right)
\end{align}
together with the fact that the basis states are eigenstates of $T^z$, $U^z$, and $V^z$.  We note that the occupation numbers change along axes that are perpendicular to the $m_t$, $m_u$, and $m_v$ axes in \figref{fig:weight_diagrams}(c).

The Rydberg blockade restricts the relevant Hilbert space to $n_r \leq 1$. We can therefore account for it by discarding all nodes in a weight diagram for which this condition is violated. Given the orientation of the $m_u$ and $m_v$ axes, \Eqref{eq:nr} implies that $n_r$ is constant along any horizontal row and increases by one for each row from top to bottom. Hence, the Rydberg blockade simply selects up to two horizontal rows from each diagram. Which rows are selected depends on the number of atoms $n$ and on the irreducible representation $(p, q)$. For the top-right corner of a weight diagram, we have $U^z = q/2$ and $V^z = (p+q)/2$, such that $n_r = (n - p - 2q)/3 = \lambda_3$ in the top row. If $\lambda_3 = 0$, the top two rows are selected, which contain zero or one Rydberg excitation. If $\lambda_3 = 1$, then only the top row is kept. Irreducible representations with larger values of $\lambda_3$ contain only states with more than one Rydberg excitation and therefore do not participate in the dynamics. Below, we will often assume that the initial state contains no Rydberg excitations. In this case, it suffices to consider partitions with $\lambda_3 = 0$.

\begin{figure}[t]
    \centering
    \includegraphics{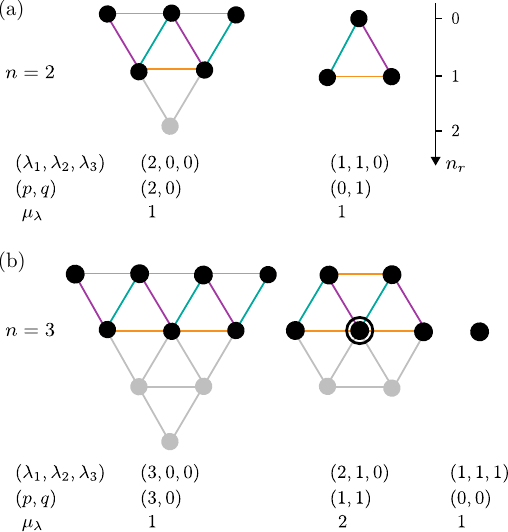}
    \caption{(a)~Irreducible representations arising for $n=2$ atoms. The axis shows the number of Rydberg atoms. States with more than one Rydberg excitation are grayed out as they violate the blockade constraint. The table shows the partition $(\lambda_1, \lambda_2, \lambda_3)$, the labels $(p, q)$, and the multiplicity $\mu_\lambda$ associated with each weight diagram. (b)~The same as (a) for $n=3$ atoms.
    }
    \label{fig:multiple_atoms}
\end{figure}

Let us illustrate these notions with the concrete examples of $n = 2$ and $n = 3$ atoms. For two atoms, there are two partitions: $(\lambda_1 , \lambda_2, \lambda_3) = (2, 0, 0)$ and $(\lambda_1 , \lambda_2, \lambda_3) = (1, 1, 0)$ corresponding to $(p, q) = (2, 0)$ and $(p, q) = (0, 1)$, respectively. The weight diagrams of these representations is shown in \figref{fig:multiple_atoms}(a), where all states with more than one Rydberg excitation have been grayed out to reflect the blockade constraint. Explicit expressions for the Hamiltonians and the states in both the computational basis and the block-diagonal basis are given in appendix \ref{app:two-atom-Hamiltonian}. Similarly, \figref{fig:multiple_atoms}(b) shows all possible partitions and the corresponding weight diagrams for $n = 3$ atoms. To account for the 27 basis states of the full Hilbert space (without blockade constraint), the partition $(\lambda_1 , \lambda_2, \lambda_3) = (2, 1, 0)$ occurs with multiplicity $\mu_\lambda = 2$. This is analogous to the fact that three spin-1/2 particles can form a system with total spin $1/2$ in two different ways.

To further elucidate the role of the multiplicity of the irreducible representations, we write the quantum state of $n$ atoms as
\begin{equation}
    \ket{\psi} = \sum_\lambda \sum_{i = 1}^{\mu_\lambda} \sum_{j = 1}^{d_\lambda} \alpha_{ij}^\lambda \ket{\lambda, i, j},
    \label{eq:state}
\end{equation}
where the sum of $\lambda$ runs over the partitions of $n$. The multiplicity $\mu_\lambda$ is accounted for by the index $i$, while the index $j$ refers to the different states of the weight diagram. The number $d_\lambda$ is the dimension of the irreducible representation restricted by the Rydberg blockade. We provide explicit expressions for both $\mu_\lambda$ and $d_\lambda$ in \appref{app:dimensions}. 

As a consequence of Schur--Weyl duality (see \appref{app:schur_weyl}), the Hamiltonian, as well as any other operator that is invariant under permutation of the atoms, decomposes into blocks that depend on the irreducible representation but not on the multiplicity index $i$. Put differently, the Hamiltonian acts nontrivially only on the index $j$ of the above basis. We may therefore express the Hamiltonian as
\begin{equation}
    \label{eq:direct_sum}
    H = \bigoplus_\lambda \mathbb{I}_{\mu_\lambda} \otimes H_\lambda ,
\end{equation}
where $\mathbb{I}_{\mu_\lambda}$ is the $\mu_\lambda \times \mu_\lambda$ identity matrix. The matrix $H_\lambda$ can be constructed by inserting the $\lambda$ representation of the operators $T^\alpha$, $U^\alpha$ and $V^\alpha$ into \eqref{eq:Hamiltonian2} and restricting the Hilbert space according to Rydberg blockade constraint (see \appref{app:matrix_elements}). Although we labeled the blocks $H_\lambda$ by the partition $\lambda = (\lambda_1, \lambda_2, \lambda_3)$, they only depend on the parameters $p$ and $q$ of the irreducible representation and are independent of $n$.

\subsection{Efficient computation}

Let us briefly summarize the insights gained above. The Hamiltonian can be brought into a block-diagonal form, where each block is associated with an irreducible representation of $\SU(3)$ labeled by two integers $p$ and $q$. The matrix elements of each block can be computed following the recipe in \appref{app:matrix_elements}. Crucially, it is not necessary to compute the unitary transformation that brings the Hamiltonian into the block-diagonal form. This is an advantage from a computational perspective because the complexity of classically constructing this transformation increases exponentially with the number of atoms~\cite{bacon2006efficient}. However, the set of quantities that can be computed without the explicit basis transformation is restricted due to the limited knowledge about the relation between the basis states in the block-diagonal form and the original tensor-product basis.

The full energy spectrum of the Hamiltonian is an example of a physically relevant quantity that can be computed without knowledge of the basis states. We will show in Section~\ref{sec:equilibration} that it is also possible to compute the number of atoms in the $\ket{0}$, $\ket{1}$, or $\ket{r}$ state after evolving an initial product state for an arbitrary time $t$. In both cases, the computational effort scales polynomially with the number of atoms, $n$. To see this, we recall that the Rydberg blockade constraint requires that $\lambda_3 \leq 1$, which implies that the number of distinct irreducible representations increases linearly with $n$. The dimension $d_\lambda$ of the unblockaded subspace in a given irreducible representation increases linearly with $p$ (see \appref{app:dimensions}), which is in turn bounded from above by $n$. Since the eigenvalues and eigenvectors of a matrix of size $m \times m$ can be determined in time $O(m^3)$~\cite{cormen2022introduction}, the computational effort to exactly diagonalize all irreducible representations is $O(n^4)$.

The polynomial scaling represents an exponential improvement over brute-force diagonalization of the Hamiltonian in the blockaded Hilbert space, which, considering the Rydberg blockade, has dimension $2^n + n 2^{n-1}$. We remark that the exponential size of the Hilbert space is accounted for by large multiplicities $\mu_\lambda$ (see \appref{app:dimensions}). The computational speedup is enabled by the fact that it suffices to diagonalize each irreducible representation once, regardless of its multiplicity.

We note that the abstract formalism in terms of irreducible representations can be useful even if knowledge of the basis states is required. We discuss such a situation in the next section, where we explore the preparation of quantum states that are invariant under permutations. Efficient computation is possible for this special case because all permutation-invariant states of $n$ atoms are part of the $(n, 0)$ representation.

\section{\label{sec:preparation} Preparation of permutation-invariant quantum states}

Having established our formalism to describe a fully blockaded system of $n$ atoms, we apply it to the preparation of arbitrary permutation-invariant quantum states. First, we note that all permutation-invariant quantum states belong to the $(p, q) = (n, 0)$ representation, which includes the states where all atoms are in the $\ket{0}$ or $\ket{1}$ state. This follows formally from the Schur--Weyl duality (see \appref{app:schur_weyl}). The basis states in the top row of the $(n, 0)$ representation are the so-called Dicke states $\ket{D^n_{n_0}}$~\cite{dicke1954coherence}, which consist of equal superposition of $n_0$ atoms in the $\ket{0}$ state and $n_1 = n - n_0$ atoms in the $\ket{1}$ state, where $n_0$ ranges from $0$ on the left to $n$ on the right of the top row. Any permutation-invariant state in the $\ket{0}$ and $\ket{1}$ basis can be decomposed into a superposition of Dicke states. We show below that one can design a pulse sequence to prepare an arbitrary permutation-invariant state starting with all atoms in the $\ket{0}$ states.

\begin{figure}
    \centering
    \includegraphics{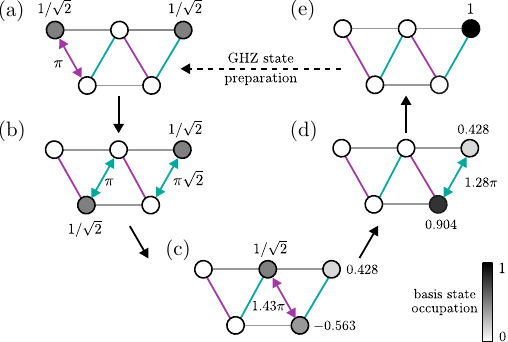}
    \caption{Preparation of the two-atom GHZ state starting from the initial state $\ket{00}$. The preparation corresponds to the time reversal of the five pulses (a)--(e). The numbers next to the arrows indicate the product $k \Omega t$ for each pulse, where $\Omega$ is the Rabi frequency for a single atom, $t$ the duration of the pulse, and $k$ the magnitude of the matrix element of the collective transition. The amplitude of each basis state (up to a global phase) is indicated by the numbers next to the nodes, which are shaded according to the corresponding occupation.}
    \label{fig:stateprep}
\end{figure}

We explain the protocol with a simple example of preparing a two-qubit Greenberger--Horne--Zeilinger (GHZ) state (or Bell state), $(\ket{00} + \ket{11}) / \sqrt{2}$, as illustrated in \figref{fig:stateprep}. The protocol closely follows the steps originally proposed in Ref.~\cite{lukin2001dipole}. The key idea is to derive a pulse sequence for the reverse process of preparing the $\ket{00}$ state from the GHZ state. The actual state preparation can then be obtained by time reversal of this pulse sequence.

Starting from the GHZ state, one can progressively move the population in the top-left node of the weight diagram ($\ket{11}$ state) to the top-right node ($\ket{00}$ state), as shown in \figref{fig:stateprep}. In each of the steps, we choose a pulse such that the population of the left-most occupied node is moved to the right. To this end, we alternate between pulses that act on the $\ket{1} \leftrightarrow \ket{r}$ and on the $\ket{0} \leftrightarrow \ket{r}$ transition. The pulse on the $\ket{1} \leftrightarrow \ket{r}$ transition can be realized using then Rabi drive $\Omega_2$ in the Hamiltonian, \eqref{eq:Hamiltonian2}. By contrast, we do not allow for a direct drive of the $\ket{0} \leftrightarrow \ket{r}$ transition in the Hamiltonian. It is possible, however, to effectively realize such a pulse by first applying a $\pi$-pulse on the $\ket{0} \leftrightarrow \ket{1}$ transition, which exchanges the $\ket{0}$ and $\ket{1}$ state of each atom. The desired pulse can then be applied to the $\ket{1} \leftrightarrow \ket{r}$ transition, before undoing the initial $\pi$-pulse. For simplicity, we leave this decomposition implied in what follows and describe the pulse sequence directly in terms of effective pulses on the $\ket{0} \leftrightarrow \ket{r}$ transition.

Owing to the Rydberg blockade, the pulses connect pairs of states between the first and second row of the weight diagram. Hence, we can determine the pulse that empties the population of the left-most state by exactly solving the dynamics of a two-level system. As we can see from \figref{fig:stateprep}(a) and (b), the first two steps are $\pi$-pulses, transferring the population from the top-left node to the bottom-left node and then to the top-middle node. A $\pi$-pulse can be implemented by setting all detunings to zero and choosing the pulse duration $t$ and the appropriate Rabi frequency $\Omega$ such that $k \Omega t = \pi$. The numerical factor $k$ accounts for the matrix element of $U^\pm$ or $V^\pm$ connecting the two states. Following the discussion of \secref{sec:review}, it is given by $k = \sqrt{2}$ for the first pulse and $k = 1$ for the second pulse.

The state-dependent matrix elements lead to a partial transfer of the population from the top-right node ($\ket{00}$ state) to the bottom-right node during the second pulse. Therefore, the third pulse is not a $\pi$-pulse. Nevertheless, we can readily determine the required pulse numerically by keeping track of matrix elements and the amplitude of each basis state, as illustrated in \figref{fig:stateprep}. The approach readily generalizes to more than $2$ atoms. The numerically calculated pulse sequences for preparing GHZ states for $n = 2$, $3$, and $4$ atoms are included in \appref{app:preparation}.

We note that the existence of the Rydberg energy level $\ket{r}$ and the blockade effect are crucial for this protocol by allowing us to isolate effective two-level systems. If we were to directly drive the $\ket{0} \leftrightarrow \ket{1}$ transition, all states on the top row would evolve together, rendering it impossible to completely move the population from the GHZ state to the state where all atoms are in the $\ket{0}$ state. This is expected as one cannot create an entangled state from a product state using only the noninteracting hyperfine states.

The above protocol straightfowardly generalizes beyond GHZ states to arbitrary permutation-invariant states, including the W state. In general, it consists of $n$ pulses on the $\ket{1} \leftrightarrow \ket{r}$ transition alternating with $n$ pulses on the $\ket{0} \leftrightarrow \ket{r}$ transition. The latter are in practice each decomposed into $3$ pulses, resulting in $4 n$ pulses in total. We highlight that the linear dependence of the number of pulses on the system size is optimal~\cite{bravyi2006lieb}. The parameters of each pulse can be computed numerically considering only effective two-level systems such that the computational cost increases only linearly with the system size.

\section{\label{sec:equilibration} Equilibration dynamics}

\subsection{Method}

We now use our formalism to study the equilibration of a permutation-invariant Rydberg system. We emphasize that we only consider the unitary dynamics of a closed system. In contrast to an open system with dissipation~\cite{yadin2022thermodynamics}, the system remains in a pure state at all times and does not converge to a mixed steady state. Equilibration in a closed system occurs as the off-diagonal elements of an observable in the energy basis dephase. At late times, the time-averaged expectation value of the observable converges to the average of its diagonal elements in the energy basis, weighted by the overlap of the initial state with the corresponding eigenstate~\cite{dalessio2016from}.

We numerically explore the time evolution of an observable for different initial states. In the examples below, we focus on the number of Rydberg excitations, $n_r$, although the approach readily applies to any observable that is invariant under permutations of the atoms. As pointed out in Section~\ref{sec:natoms}, any such observable $A$ can be written as
\begin{equation}
    A = \bigoplus_\lambda \mathbb{I}_{\mu_\lambda} \otimes A_\lambda,
\end{equation}
where the direct sum runs over the irreducible representations of $\SU(3)$ labeled by $\lambda$. The matrices $A_\lambda$ describe the action of $A$ on the vector space associated with the irreducible representation. In the case of $n_r$, these irreducible blocks can be readily constructed using \eqref{eq:nr}.

Given a state $\ket{\psi}$, the expectation value of $A$ may be expressed as
\begin{equation}
    \label{eq:projection}
    \braket{\psi | A | \psi} = \sum_{\lambda} \Tr ( A_\lambda \rho_\lambda ),
\end{equation}
where $\rho_\lambda$ is the (unnormalized) state obtained by projecting $\ket{\psi} \bra{\psi}$ onto the irreducible representation labeled by $\lambda$ and tracing out its multiplicity. Using the notation of \eqref{eq:state}, $\rho_\lambda$ can be written explicitly as
\begin{equation}
    \rho_\lambda = \sum_{i = 1}^{\mu_\lambda} \sum_{j = 1}^{d_\lambda} \sum_{j' = 1}^{d_\lambda} \alpha_{ij}^\lambda \alpha_{ij'}^{\lambda *} \ket{\lambda, j} \bra{\lambda, j'}.
\end{equation}
Here, the states $\{ \ket{\lambda, j} \}$ form an orthonormal basis of the $d_\lambda$-dimensional vector space associated with the irreducible representation. Since the projection commutes with the Hamiltonian, we can apply the time evolution to each $\rho_\lambda$ separately. Given the initial state $\rho_\lambda(0)$, we can classically compute the time-evolved state $\rho_\lambda(t) = e^{- i H_\lambda t} \rho_\lambda(0) e^{i H_\lambda t}$ at a computational cost that increases polynomially with the system size. Hence, it is possible to efficiently determine the time-dependent expectation value of permutationally invariant observables.

We focus on initial states with a well-defined number of atoms in the $\ket{0}$ and $\ket{1}$ states and no atom in the Rydberg state, $\ket{r}$. According to Eqs.~(\ref{eq:n0})--(\ref{eq:nr}), such states are eigenstates of $T^z$, $U^z$, and $V^z$. The respective eigenvalues uniquely specify the node in the weight diagram of each irreducible representation. Moreover, since $n_r = 0$, only irreducible representations with $\lambda_3 = 0$ contribute and the node lies in the top row. Because all nodes in the top row host nondegenerate states, the occupation numbers specify a unique initial state for each irreducible representation. We denote this state by $\ket{\lambda, n_0, n_1}$, where $n_0$ and $n_1$ are the number of atoms in the $\ket{0}$ and $\ket{1}$ states, respectively. Taking this state to be normalized, we have
\begin{equation}
    \rho_\lambda(0) = w_\lambda \ket{\lambda, n_0, n_1} \bra{\lambda, n_0, n_1}
\end{equation}
where $w_\lambda$ is the probability that the system occupies the irreducible representation labeled by $\lambda$. We show in \appref{app:projector} that 
\begin{equation}
    w_\lambda = (p + 1) \frac{n_0!n_1!}{q!(p + q + 1)!}
    \label{eq:weight}
\end{equation}
if $\lambda_3 = 0$ and $|n_0-n_1| \leq p \leq n$, and $w_\lambda = 0$ otherwise. The representations $A_\lambda$ with $A = n_r$ are diagonal matrices with entries $0$ for the states corresponding to the top row of the weight diagram and $1$ for the ones in the bottom row.

\begin{figure*}
    \centering
    \includegraphics{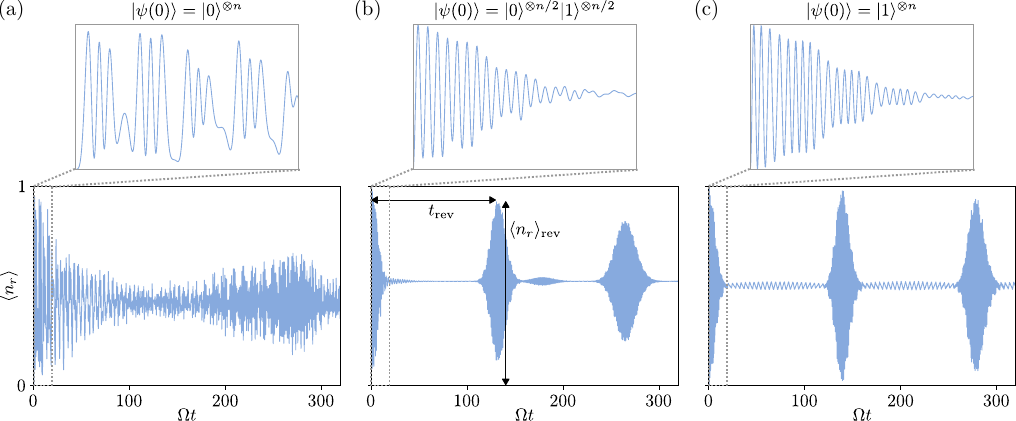}
    \caption{Expected number of Rydberg excitations in a fully-blockaded ensemble of $n = 100$ atoms. The dynamics are shown for three different initial states, indicated at the top of each panel. The system is simultaneously driven by two external fields with Rabi frequencies $\Omega_1 = \Omega_2 = \Omega$ and detunings $\Delta_1 = \Omega$ and $\Delta_2 = 0$ (see \figref{fig:rydbergblockade}). The plots in the top row show the rapid oscillations of the Rydberg population for times between $\Omega t = 0$ and $\Omega t = 20$.}
    \label{fig:fig5}
\end{figure*}

\subsection{Numerical results}

We next numerically study the dynamics of large systems with different initial states. \Figref{fig:fig5} shows the results for $n = 100$ atoms with initial product states of the form $\ket{\psi(0)} = \ket{0}^{\otimes n_0} \ket{1}^{\otimes n_1}$. We remark that due to permutation invariance, it is irrelevant which of the atoms are in the $\ket{0}$ state and which in the $\ket{1}$ state. The system is subject to simultaneous driving of the $\ket{0} \leftrightarrow \ket{1}$ and the $\ket{1} \leftrightarrow \ket{r}$ transitions with equal Rabi frequencies, $\Omega_1 = \Omega_2 = \Omega$. The field driving the hyperfine transition is detuned by $\Delta_1 = \Omega$, whereas the Rydberg drive is resonant, $\Delta_2 = 0$. We note that only the $(p, q) = (n, 0)$ representation is relevant for Figs.~\ref{fig:fig5}(a) and (c), where all atoms occupy the same state, rendering the state invariant under permutations. This is not the case in \figref{fig:fig5}(b), for which we have to sum over all irreducible representations satisfying $\lambda_3 = 0$.

For all three initial states in \figref{fig:fig5}, the Rydberg population oscillates rapidly. We note that the typical period of the oscillation is much shorter than $2 \pi / \Omega_2$, which is the duration of a $2 \pi$ pulse between the $\ket{1}$ and $\ket{r}$ states of a single atom. This can be explained by the enhanced Rabi frequency experienced by states with more than one atom in the $\ket{1}$ state. The Rabi frequency experienced by the $\ket{1}^{\otimes n}$ state is given by $\sqrt{n} \Omega_2$, which shortens the period by a factor $1 / \sqrt{n}$. Other states experience a weaker enhancement, which depends on both the number of atoms in the $\ket{0}$ and $\ket{1}$ states as well as the irreducible representation. Following the construction of matrix elements in Section~\ref{sec:setup} and \appref{app:matrix_elements}, we find that the largest enhancement factor in the $(p, q)$ representation is $\sqrt{(n + p) / 2}$. Hence an enhancement factor proportional to $\sqrt{n}$ is expected independent of the initial state.

In addition to the fast oscillation, the dynamics exhibit a much slower modulation of the oscillation amplitude. When the atoms all start in the state $\ket{0}$ (\figref{fig:fig5}(a)), the fast oscillations as well as the slow modulation possess little structure. By contrast, when half or all atoms start in the $\ket{1}$ state (Figs.~\ref{fig:fig5}(b) and (c)), the system undergoes damped oscillations with strong revivals after a time of more than $100 / \Omega$. Revivals in such a large system are surprising and are often indicators of symmetries such as the permutational invariance in the present setting.

We characterize the revivals by determining the time and strength of the revival. We accomplish this numerically by sorting the evolution times according to their corresponding Rydberg population. The largest Rydberg populations occur at short times. As the Rydberg population decreases, the sorted evolution time jumps up when the highest point of a revival is reached. Because the strength of the revivals is monotonically decreasing, this jump reliably identifies the first revival. We refer to the time and Rydberg population at this point as the revival time $t_\mathrm{rev}$ and the revival strength $\langle n_r \rangle_\mathrm{rev}$, respectively (see \figref{fig:fig5}(b)). We plot both quantities in \Figref{fig:fig6} as a function of the number of atoms for the two initial states that exhibit pronounced revivals.

Figure~\ref{fig:fig6}(a) shows that the time of the first revival increases with $\sqrt{n}$. The functional dependence of the revival strength on $n$ is not evident from the data in \figref{fig:fig6}(b), but the revival strength remains above $0.9$ even for $n = 300$ atoms. This indicates that the first revival stays pronounced in large systems even as it shifts to later times. The strength of subsequent revivals, however, decreases and data at very late times show that they eventually disappear (see \appref{app:late}). 

\subsection{Spin model}

To shed light on the physical origin of the revivals, we consider the dynamics of the initial state $\ket{1}^{\otimes n}$ in more detail. In this case, the evolution is restricted to the irreducible representation $(p, q) = (n, 0)$, whose weight diagram is shown in \figref{fig:fig7}(a). The top row of this diagram, where no atom is in the Rydberg state, may be viewed as the states of a spin with total angular momentum quantum number $s = n/2$. Similarly, the bottom row, where exactly one atom is in the Rydberg state, represents a spin $s = (n-1)/2$. In the limit of large $n$, we may ignore the difference between these spins. We formalize this idea by adding an unphysical state to the bottom row. This additional state will have little impact on the dynamics provided its amplitude remains negligible throughout the evolution. We expect this to be indeed the case because the initial state is situated at the top-left corner of the weight diagram and the detuning $\Delta_1 = \Omega$ creates an energy difference of order $n \Delta_1$ between opposite ends of the diagram such that occupying the added state is energetically unfavorable.

The addition of the unphysical state leads to a great simplification because both rows now have $n + 1$ states. This allows us to factorize the Hilbert space into a spin $s = n/2$ and a two-level system whose states $\ket{0}$ and $\ket{1}$ correspond to the number of atoms in the Rydberg state. As a further approximation, we modify the matrix elements of $T^{x,y,z}$ in the bottom row, which originally correspond to those of a spin $s = (n-1)/2$. In the limit of large $n$, it is justified to replace the matrix elements by those of a spin $s = n/2$. The spin-model Hamiltonian resulting from these approximations can be compactly expressed as
\begin{align}
    \label{eq:spin_model}
    H_\text{SM} &= \left( \Omega_1 S^x + \Delta_1 S^z \right) \otimes \mathbb{I}_2 \nonumber \\
    & \quad + \frac{\Delta_2}{2}  \mathbb{I}_{2 s + 1} \otimes \sigma^z + \frac{\Omega_2}{2} \sqrt{s \mathbb{I}_{2 s + 1} - S^z} \otimes \sigma^x,
\end{align}
where $\sigma^{x,y,z}$ are Pauli matrices and $S^{x,y,z}$ are the standard spin operators with total angular momentum $s = n/2$. For simplicity, we assumed that $\Omega_1$ and $\Omega_2$ are real. The operators $S^{x,y,z}$ couple the same states as $T^{x,y,z}$ with the above modification of the matrix elements and the additional, unphysical state. The term $\sqrt{s \mathbb{I}_{2 s + 1} - S^z}$ accounts for the matrix elements of the $U^\pm$ operators in the original Hamiltonian (see \figref{fig:fig7}(a)). We note that the Pauli matrices are related to the occupation of the Rydberg state by $n_r = \mathbb{I}_{2 s + 1} \otimes (\mathbb{I}_2 - \sigma^z ) / 2$.

\begin{figure}[t]
    \centering
    \includegraphics{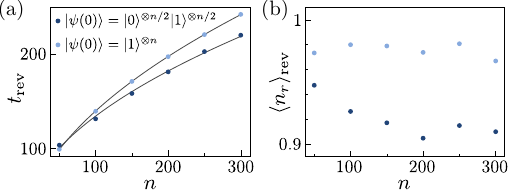}
    \caption{(a) Revival time $t_{\text{rev}}$ and (b) revival strength $\langle n_r \rangle_\mathrm{rev}$ as a function of the number of atoms, $n$. Both quantities are determined by locating the largest value of $\langle n_r \rangle$ at the first revival. The Hamiltonian parameters are the same as in \figref{fig:fig5}. The solid curves in (a) are fits to the function $a \sqrt{n} + b$.
    }
    \label{fig:fig6}
\end{figure}

In the special case $\Delta_2 = 0$, the states $\ket{\pm} = (\ket{0} \pm \ket{1})/\sqrt{2}$ of the two-level system are decoupled. It is therefore convenient to express the state of the system as
\begin{equation}
    \ket{\psi(t)} = \ket{\phi_+(t)} \ket{+} + \ket{\phi_-(t)} \ket{-}.
    \label{eq:ansatz}
\end{equation}
The states $\ket{\phi_\pm(t)}$ evolve under the Hamiltonian
\begin{equation}
    H_\text{SM}^{ \pm} = \Omega_1 S^x + \Delta_1 S^z \pm \Omega_2 \sqrt{s \mathbb{I}_{2s+1} - S^z} / 2.
    \label{eq:sm}
\end{equation}
In terms of this ansatz, the number of atoms in the Rydberg state is given by
\begin{equation}
    \langle n_r(t) \rangle = \frac{1}{2} - \, \mathrm{Re} \braket{\phi_+(t) | \phi_-(t)}.
\end{equation}
This implies that $1/2 - |\braket{\phi_+(t) | \phi_-(t)}| \leq \langle n_r(t) \rangle \leq 1/2 + |\braket{\phi_+(t) | \phi_-(t)}|$. The absolute value of the overlap between $\ket{\phi_+(t)}$ and $\ket{\phi_-(t)}$ hence determines the envelope of the Rydberg population, whereas the relative phase is responsible for the fast oscillations.

\begin{figure}
    \centering
    \includegraphics[width=\linewidth]{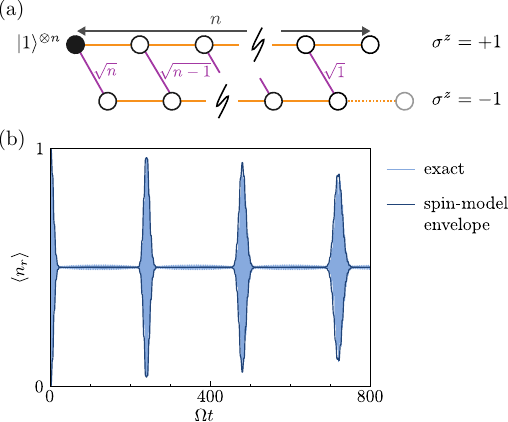}
    \caption{(a)~Illustration of the spin model approximation to the $(n,0)$ representation. By adding an unphysical state (gray circle connected to dashed line), the Hilbert space can be viewed as the tensor product of a spin $s = n/2$ and a two-level system $\sigma$. The colored lines indicate the transitions that are being driven. The purple numbers show the matrix elements for the $\ket{1} \leftrightarrow \ket{r}$ transition. (b)~Revivals of an ensemble of $n = 300$ atoms initialized in the state $\ket{1}^{\otimes n}$. The Hamiltonian parameters are the same as in \figref{fig:fig5}. The gray curve shows the exact values of the expectation value of the number of Rydberg excitations. The blue envelope was obtained from the approximate spin model.}
    \label{fig:fig7}
\end{figure}

To verify the validity of the spin model, we plot in \figref{fig:fig7}(b) the exact dynamics of $n = 300$ atoms together with the spin-model envelope $1/2 \pm |\braket{\phi_+(t) | \phi_-(t)}|$. The atoms are initialized in the state $\ket{1}^{\otimes n}$ and the Hamiltonian parameters are the same as in Figs.~\ref{fig:fig5} and~\ref{fig:fig6}. There is excellent agreement between the exact evolution and the envelope. We highlight that the spin model also captures the fast oscillations, which have been omitted from the figure for clarity.

The spin model enables us to explain the characteristics of the revivals displayed in \figref{fig:fig6}. In fact, it is sufficient to work within a semiclassical approximation, which is valid in the limit of large $s$. We write $S^z = \langle S^z(t) \rangle + \delta S^z(t)$ and expand $\sqrt{s \mathbb{I}_{2 s + 1} - S^z}$ to linear order in $\delta S^z(t)$. This yields
\begin{equation}
    \sqrt{s \mathbb{I}_{2 s + 1} - S^z} \approx \sqrt{s - \langle S^z(t) \rangle}  - \frac{S^z - \langle S^z(t) \rangle}{2 \sqrt{s - \langle S^z(t) \rangle}}.
    \label{eq:large_spin}
\end{equation}
The expansion is justified if the fluctuation $\delta S^z(t)$ is small compared to $s - \langle S^z(t) \rangle$. This is indeed the case for sufficiently large detuning $\Delta_1$: Since $\langle S^z(0) \rangle = -s$, the detuning ensures that $s - \langle S^z(t) \rangle$ is of order $s$ at all times. In comparison, the fluctuations in the initial state are smaller by a factor $1/\sqrt{s}$ according to $\sqrt{\langle \delta S^z(0)^2 \rangle} \sim \sqrt{s}$. We note that the uncertainty in the spin vector remains constant within the approximations of \eqref{eq:large_spin}. Although higher-order corrections may cause the fluctuations to grow, we expect those terms to only become relevant at very late times.

By substituting \eqref{eq:large_spin} into the expression for the spin model Hamiltonian in \eqref{eq:sm}, we find that the states $\ket{\phi_\pm(t)}$ evolve according to
\begin{align}
    H_\text{SM}^{ \pm} = &\pm \frac{\Omega_2}{2} \left[ \sqrt{s - \langle S^z(t) \rangle}  + \frac{\langle S^z(t) \rangle}{2 \sqrt{s - \langle S^z(t) \rangle}} \right] \mathbb{I}_{2 s + 1} \nonumber \\
    & + \Omega_1 S^x + \left[ \Delta_1 \mp \frac{\Omega_2}{4 \sqrt{s - \langle S^z(t) \rangle}}\right] S^z.
\end{align}
The term proportional to the identity leads to a phase difference between the states and thus causes oscillations in the Rydberg population with a frequency of order $\sqrt{n} \Omega_2$. The second row of the Hamiltonian describes an effective magnetic field around which the states $\ket{\phi_\pm(t)}$ precess. Since there are no nonlinear terms, the states remain spin-coherent states at all times and can be simply described by the direction of their spin vector. The state overlap can be expressed as $| \braket{\phi_+(t) | \phi_-(t)} | = \cos^{2 s} (\theta / 2)$, where $\theta$ is the angle between the two respective spin vectors~\cite{radcliffe1971some}. The magnitude of the magnetic fields experienced by the two states differs by a term proportional to $\Omega_2/\sqrt{s - \langle S^z(t) \rangle}$. Hence, the two spin vectors separate and rephase over a time of order $\sqrt{s}$, which explains the revival timescale $t_\mathrm{rev} \sim \sqrt{n}$ observed in \figref{fig:fig6}(a). The revival is imperfect because the rotation axes fluctuate such that the spins do not return to the exact initial direction but are displaced by an angle of order $\Delta \theta \sim 1/\sqrt{s}$. We expect that the strength of the first revival, $\langle n_r \rangle_\mathrm{rev}$, is independent of the system size since $\cos^{2 s} (c / \sqrt{s})$ tends to $e^{-c^2/2}$ in the limit of large $s$ for any constant $c$. This is consistent with the weak dependence on the system size observed in \figref{fig:fig6}(b).

We have shown that the spin model captures the salient features of the revivals when the system starts in the state $\ket{1}^{\otimes n}$. The system fails to equilibrate quickly because it is well approximated by a linear spin model. As pointed out above, the revivals decay over time, which is also true within the spin model. The approximations of the spin model, however, render it inapplicable at very late times, as can be seen in \figref{fig:figs1} of \appref{app:late}.

The spin model can be adapted to the initial state $\ket{0}^{\otimes n}$ (see \appref{app:model}). The additional, unphysical state must be added to the left end of the weight diagram to ensure its occupation remains small. However, the resulting Hamiltonian is more complicated than \eqref{eq:spin_model} because driving the $\ket{1} \leftrightarrow \ket{r}$ transition results in a term that simultaneously changes the large spin $s$ and flips the two-level system. The dynamics can therefore not be separated as in \eqref{eq:ansatz}. As a consequence, the large spin and the two-level system do not periodically disentangle, which explains the absence of revivals for this initial state. For the initial state $\ket{0}^{\otimes n/2} \ket{1}^{\otimes n/2}$, a similar analysis is more involved because it requires summing over multiple irreducible representations. Moreover, a two-level system coupled to a large spin is insufficient to capture the two states per node that appear in the second row of many representations. The pronounced revivals for this initial state are thus surprising and a more refined model is required to explain their physical origin.

\section{Summary and Outlook\label{sec:summary}}

In summary, we presented a general formalism to efficiently compute the quantum dynamics of an ensemble of three-level systems evolving under a permutation-invariant Hamiltonian. The high degree of symmetry allows us to numerically simulate ensembles of $100$s of atoms that interact via the Rydberg blockade mechanism. We applied the formalism to derive sequences of at most $4 n$ pulses to prepare arbitrary symmetric states of $n$ atoms, including important entangled states such as the W state and the GHZ state. In addition, we studied the dynamics of an ensemble after a quench and uncovered a regime of parameters in which the system approaches equilibrium surprisingly slowly despite the strong nonlinearity due to the Rydberg blockade. We explain this slow equilibration and the appearance of revivals in terms of an effective spin model for permutation-invariant initial states. The physical origin of the revivals for other initial states will be the subject of future research.

Our formalism generalizes the notion of Dicke states from two-level to three-level systems. The approach is general and can accommodate other forms of interaction besides Rydberg blockade, provided the Hamiltonian remains invariant under permutations. It is possible to generalize our results beyond three-level systems to arbitrary $d$-level systems as the underlying Schur--Weyl duality also applies to the group $\SU(d)$. Future work may further extend the formalism to open systems governed by a permutationally invariant Lindblad operator~\cite{shammah2018open,yadin2022thermodynamics}.

\begin{acknowledgments}
    We thank Nate Gemelke, Daniel Malz, and Zhi-Yuan Wei for helpful discussions. This work is supported by DARPA-STTR award (Award No.~140D0422C0035).
\end{acknowledgments}

\bibliography{refs}

\clearpage
\appendix

\section{Schur--Weyl duality\label{app:schur_weyl}}

Schur--Weyl duality formalizes an intimate relation between representations of the symmetric group $\mathrm{S}_n$ and and the general linear group $\mathrm{GL}_d$. We consider a system formed by $n$ qudits with local Hilbert space dimension $d$. The duality states that the corresponding Hilbert space decomposes according to
\begin{equation}
  \left( \mathbb{C}^d \right)^{\otimes n} \cong \bigoplus_\lambda P_\lambda \otimes Q_\lambda,
  \label{eq:schur_weyl}
\end{equation}
where the sum runs over all partitions of $n$ into $d$ nonnegative integers. Each partition $\lambda$ labels an irreducible representation of $\mathrm{S}_n$ and of $\mathrm{GL}_d$.

The duality further states that representations of $\mathrm{S}_n$ and $\mathrm{GL}_d$ act nontrivially only on the respective vector spaces $P_\lambda$ and $Q_\lambda$. Concretely, the representation $M(\pi)$ that permutes the qudits according to some $\pi \in \mathrm{S}_n$ can be written as
\begin{equation} 
    M(\pi) = \bigoplus_\lambda M_\lambda(\pi) \otimes \mathbb{I}_{\dim Q_\lambda},
\end{equation}
where $M_\lambda(\pi)$ is an irreducible representation of the symmetric group. Similarly, given the representation $N(U)$ of $U \in \mathrm{GL}_d$ that acts on $\left( \mathbb{C}^{d} \right)^{\otimes n}$ as $U^{\otimes n}$, we have
\begin{equation}
  N(U) = \bigoplus_\lambda \mathbb{I}_{\dim P_\lambda} \otimes N_\lambda(U).
\end{equation}
Here, $N_\lambda(U)$ is an irreducible representations of $\mathrm{GL}_d$.

In the main text, we are concerned with the special unitary group $\SU(d)$ instead of $\mathrm{GL}_d$. Schur--Weyl duality continues to apply since $\SU(d)$ is a subgroup of $\mathrm{GL}_d$. The duality further constrains the form of operators that commute with all permutations. According Schur's lemma, any such operator $O$ must take the form
\begin{equation}
    O = \bigoplus_\lambda \mathbb{I}_{\dim P_\lambda} \otimes O_\lambda.
\end{equation}

We further note that the partition $\lambda$ with $\lambda_1 = n$ and all other $\lambda_k = 0$ ($k = 2, 3, \ldots, d$) describes the trivial representation $\rho_\lambda(\pi) = 1$ of the symmetric group. Hence, the contribution of this partition to the direct sum of \eqref{eq:schur_weyl} corresponds to the subspace of states that are invariant under permutations.

\section{\texorpdfstring{Irreducible representations of $\su(3)$}{Irreducible representations of su(3)}}

\subsection{Dimensions \label{app:dimensions}}

Here, we give expressions for the dimensions of the irreducible representations $P_\lambda$ and $Q_\lambda$ of the symmetric group $\mathrm{S}_n$ and the special unitary group $\SU(3)$, respectively. The representations are labeled by $\lambda = ( \lambda_1, \lambda_2, \lambda_3)$ which partition $n$ such that $\lambda_1 + \lambda_2 + \lambda_3 = n$. For concreteness, we order the three integers as $\lambda_1 \geq \lambda_2 \geq \lambda_3 \geq 0$. For conciseness, we also use the quantities $p = \lambda_1 - \lambda_2$ and $q = \lambda_2 - \lambda_3$.

In the main text, we refer to the dimension of the irreducible representation $P_\lambda$ of $\mathrm{S}_n$ as the multiplicity $\mu_\lambda$. It can be computed using the hook formula (see chapter 2.8 of reference~\cite{sternberg1994group}), which gives
\begin{equation}
    \mu_\lambda = \dim P_\lambda = n! \frac{(p + q + 2) (p + 1) (q + 1)}{(\lambda_1 + 2)! (\lambda_2 + 1)! \lambda_3!} .
\end{equation}

The dimension of the irreducible representation $Q_\lambda$ of $\su(3)$ can be obtained by directly counting of the states in the weight diagram. We find
\begin{equation}
    \dim Q_\lambda = \frac{1}{2} (p + q + 2) (p + 1) (q + 1) .
\end{equation}
We note that in the presence of Rydberg blockade, the states with more than one Rydberg excitation do not participate in the dynamics. According to the discussion in the main text, the number of Rydberg excitations is bounded by $n_r \geq \lambda_3$. It follows that the subspace $\tilde Q_\lambda$ of $Q_\lambda$ with at most one Rydberg excitation has dimension
\begin{equation}
    d_\lambda = 
    \begin{cases}
        3 ( p + 1 ) - (p + 2) \delta_{q,0} & \text{if } \lambda_3 = 0\\
        p + 1 & \text{if } \lambda_3 = 1\\
        0 & \text{otherwise}
    \end{cases} .
\end{equation}

\subsection{Commutation relations \label{app:commutators}}

The generators of the Lie algebra $\su(3)$ form three $\su(2)$ subalgebras with the commutation relations
\begin{align}
  \comm{T^\pm}{T^z} &= \pm T^\pm, \qquad
  \comm{T^+}{T^-} = 2 T^z,\\
  \comm{U^\pm}{U^z} &= \pm U^\pm, \qquad
  \comm{U^+}{U^-} = 2 U^z,\\
  \comm{V^\pm}{V^z} &= \pm V^\pm, \qquad
  \comm{V^+}{V^-} = 2 V^z.
\end{align}
The $\su(3)$ algebra is fully defined by these commutation relations together with
\begin{align}   \comm{T^+}{V^-} &= - U^- , \\
  \comm{T^+}{U^+} &= V^+ , \\
  \comm{U^+}{V^-} &= T^-\\ 
  2 \comm{T^z}{U^\pm} &= \mp U^\pm, \\
  2 \comm{T^z}{V^\pm} &= \pm V^\pm\\
  \comm{T^+}{V^+} = \comm{T^+}{U^-} &= \comm{U^+}{V^+} = 0 .
\end{align}
Further commutation relations can be derived by nesting the above equations and taking the Hermitian conjugates.

\subsection{Matrix elements \label{app:matrix_elements}}

In this section, we describe how to explicitly construct irreducible representations of the $\su(3)$ algebra. 
We label the states in a given diagram corresponding to the irreducible representation $(p, q)$ by $\ket{r; t, m_t}$. Here, $r$ is a non-negative integer assigned to each row, with $r = 0$ corresponding to the top row. The value of $r$ increases by one each time that we move down by one row. The other two labels are angular momentum eigenvalues such that
\begin{align}
    T^2 \ket{r; t, m_t} &= t ( t + 1 ) \ket{r; t, m_t}, \\
    T^z \ket{r; t, m_t} &= m_t \ket{r; t, m_t},
\end{align}
where $T^2 = (T^x)^2 + (T^y)^2 + (T^z)^2$. We may view $m_t$ as a horizontal coordinate in the weight diagram. Its values are restricted to $\{ - t, -t + 1, \ldots, t \}$. The total angular momentum quantum number $t$ serves to resolve different states on multiply occupied nodes of the weight diagram. Since the operators $T^\alpha$ form a $\su(2)$ subalgebra, we further have
\begin{align}
    T^\pm &\ket{r; t, m_t} \nonumber\\
    & = \sqrt{t(t+1) - m_t (m_t \pm 1)} \ket{r; t, m_t \pm 1}.
\end{align}

We further note that the states in the weight diagram are simultaneous eigenstates of $U^z$ and $V^z$. Using the relations $U^z - V^z + T^z = 0$ and $r = ( p + 2 q - 2 U^z - 2 V^z )$, we find
\begin{align}
    U^z \ket{r; t, m_t} &= \frac{1}{4} (p + 2 q - 3 r - 2 m_t) \ket{r; t, m_t},\\
    V^z \ket{r; t, m_t} &= \frac{1}{4} (p + 2 q - 3 r + 2 m_t) \ket{r; t, m_t}.
\end{align}

It remains to determine the matrix elements of $U^\pm$ and $V^\pm$. These matrix elements are more complicated because they couple states with different values of $t$. As we are only interested in matrix elements between states with at most one Rydberg excitation, we will restrict the remaining discussion to the top two rows of any weight diagram, which limits the range of values of $t$. The top row is always singly occupied with $t = p/2$. To determine the allowed values for $t$ in the second row, we distinguish between four cases.
\begin{description}
    \item[$p = q = 0$] The irreducible representation is one-dimensional and there is only a single state, for which $t = 0$.
    \item[$p = 0, q \neq 0$] All nodes in the second row are singly occupied with $t = 1/2$.
    \item[$p \neq 0, q = 0$] All nodes in the second row are singly occupied with $t = (p-1)/2$.
    \item[$p \neq 0, q \neq 0$] The second row supports states with $t = (p + 1)/2$ and $t = (p-1)/2$. The two edge nodes remain singly occupied, while all inner nodes are doubly occupied.
\end{description}

The matrix elements can now be constructed from the commutation relations of the Lie algebra~\cite{greiner1989quantum}. Given an integer $0 \leq k \leq p/2$, they can be concisely written as
\begin{align}
    U^- & \ket{0; \frac{p\vphantom{1}}{2}, \frac{p}{2} - k} = \sqrt{\frac{q(p-k + 1)}{p+1}} \ket{1; \frac{p+1}{2}, \frac{p+1}{2} - k} \nonumber\\
    & +  \sqrt{\frac{k(p+q+1)}{p+1}}\ket{1; \frac{p-1}{2}, \frac{p+1}{2} - k},\\
    V^- &\ket{0; \frac{p\vphantom{1}}{2}, \frac{p}{2} - k} = - \sqrt{\frac{q(k+1)}{p+1}} \ket{1; \frac{p+1}{2}, \frac{p-1}{2} - k} \nonumber\\
    & + \sqrt{\frac{(p - k)(p+q+1)}{p+1}} \ket{1; \frac{p-1}{2}, \frac{p-1}{2} - k}.
\end{align}
The matrix elements of $U^+$ and $V^+$ follow by taking the Hermitian conjugate. The expressions hold in all four of the above cases. When $p = 0$ or $q = 0$, some of the coefficients on the right-hand side vanish such that only states with either $t = (p+1)/2$ or $t = (p-1)/2$ are present.

\subsection{\texorpdfstring{Hamiltonian for $n = 2$ atoms}{Hamiltonian for n = 2 atoms}\label{app:two-atom-Hamiltonian}}

In this appendix, we give explicit expressions for the Hamiltonian for $n = 2$ atoms, both in the original basis and in the block-diagonal form. Taking into account the Rydberg blockade, the Hilbert space is spanned by the eight states $\{ \ket{00}, \ket{01}, \ket{0r}, \ket{10}, \ket{11}, \ket{1r}, \ket{r0}, \ket{r1} \}$. Using this ordering of the states, the Hamiltonian in \eqref{eq:Hamiltonian1} has the matrix representation
\begin{widetext}
\begin{equation}
    H =
    \begin{pmatrix}
        0 & \Omega_1 / 2 & 0 & \Omega_1 / 2 & 0 & 0 & 0 & 0\\
        \Omega_1 / 2 & -\Delta_1 & \Omega_2 / 2 & 0 & \Omega_1 / 2 & 0 & 0 & 0\\
        0 & \Omega_2 / 2 & -\Delta_1 - \Delta_2 & 0 & 0 & \Omega_1 / 2 & 0 & 0\\
        \Omega_1 / 2 & 0 & 0 & -\Delta_1 & \Omega_1 / 2 & 0 & \Omega_2 / 2 & 0\\
        0 & \Omega_1 / 2 & 0 & \Omega_1 / 2 & -2\Delta_1 & \Omega_2 / 2 & 0 & \Omega_2 / 2\\
        0 & 0 & \Omega_1 / 2 & 0 & \Omega_2 / 2 & -2 \Delta_1 - \Delta_2 & 0 & 0\\
        0 & 0 & 0 & \Omega_2 / 2 & 0 & 0 & -\Delta_1 - \Delta_2 & \Omega_1 / 2\\
        0 & 0 & 0 & 0 & \Omega_2 / 2 & 0 & \Omega_1 / 2 & -2 \Delta_1 - \Delta_2
    \end{pmatrix}.
\end{equation}
\end{widetext}

Following the discussion of \secref{sec:setup}, this Hamiltonian can be brought into block-diagonal form with two blocks corresponding to the partitions $\lambda = (2, 0, 0)$ and  $\lambda = (1, 1, 0)$. The blocks are explicitly given by
\begin{widetext}
\begin{equation}
    H_{(2,0)} =
    \begin{pmatrix}
        0 & \Omega_1 / \sqrt{2} & 0 & 0 & 0\\
        \Omega_1 / \sqrt{2} & -\Delta_1 & \Omega_1 / \sqrt{2} & \Omega_2 / 2 & 0\\
        0 & \Omega_1 / \sqrt{2} & -2 \Delta_1 & 0 & \Omega_2 / \sqrt{2}\\
        0 & \Omega_2 / 2 & 0 & -\Delta_1 - \Delta_2 & \Omega_1 / 2\\
        0 & 0 & \Omega_2 / \sqrt{2} & \Omega_1 / 2 & -2 \Delta_1 - \Delta_2
    \end{pmatrix},
    \quad   
    H_{(0,1)} =
    \begin{pmatrix}
        - \Delta_1 & \Omega_2 / 2 & 0\\
        \Omega_2 / 2 & -\Delta_1 - \Delta_2 & \Omega_1 / 2\\
        0 & \Omega_1 / 2 & -2 \Delta_1 - \Delta_2
    \end{pmatrix}.
\end{equation}
\end{widetext}
Here, we labeled the blocks by $(p, q)$ instead of the partition $\lambda$ to highlight that their form also applies to other values of $n$.

One can explicitly check that $H_{(2,0)}$ and $H_{(1,0)}$ correspond to the matrix elements of $H$ in the subspaces respectively spanned by
\begin{align}
    &\ket{0; 1, 1} = \ket{0 0}\\
    &\ket{0; 1, 0} = \frac{1}{\sqrt{2}} (\ket{0 1} + \ket{1 0})\\
    &\ket{0; 1, -1} = \ket{1 1}\\
    &\ket{1; 1/2, 1/2} = \frac{1}{\sqrt{2}} (\ket{0 r} + \ket{r 0})\\
    &\ket{1; 1/2, -1/2} = \frac{1}{\sqrt{2}} (\ket{1 r} + \ket{r 1})
\end{align}
and
\begin{align}
    &\ket{0; 0, 0} = \frac{1}{\sqrt{2}} (\ket{0 1} - \ket{1 0})\\
    &\ket{1; 1/2, 1/2} = \frac{1}{\sqrt{2}} (\ket{0 r} - \ket{r 0})\\
    &\ket{1; 1/2, -1/2} = \frac{1}{\sqrt{2}} (\ket{1 r} - \ket{r 1}).
\end{align}
These are the basis states associated with the nodes of the weight diagrams labeled as in \appref{app:matrix_elements}. We observe that the states in the $(p, q) = (2, 0)$ irreducible representation are invariant under permutation. The states in the $(p, q) = (0, 1)$ representations change sign under permutation. The states can be constructed by first constructing the eigenstates of total angular momentum of two atoms with two levels $\ket{0}$ and $\ket{1}$. The remaining states can subsequently be obtained by applying the operators $U^-$ and $V^-$.

\subsection{Projector\label{app:projector}}

In this appendix, we derive a closed-form expression for the overlap of a given state with the $(p, q)$ representation. We consider initial product states of the form $\ket{s_1} \ket{s_2} \cdots \ket{s_n}$, where $s_i \in \{ 0, 1 \}$ for all $i \in \{ 1, 2, \ldots, n \}$. Since there are no Rybderg excitations for this initial state, the overlap with the $(p, q)$ representation is equal to the probability that the product state carries total spin quantum number $t = p/2$ with respect to the operator $T^2$.

Given a finite-dimensional representation $\rho$ of a compact group $G$, the projector, $P_\lambda$, onto an irreducible representation $\rho_\lambda$ of dimension $d_\lambda$ can be written as (see chapter 5 of reference~\cite{cornwell1997group})
\begin{align}
    P_\lambda = d_\lambda \int \di U \, \chi_\lambda^*(U) \rho(U).
\end{align}
Here, the integral runs over the Haar measure on $G$ and $\chi_\lambda(U) = \tr [ \rho_\lambda ( U ) ]$ is the character of the group element in the irreducible representation.

Since the Rydberg level plays no role, the group of interest is $\SU(2)$. The elements of $\SU(2)$ can be described by the rotation axis $\vec{\hat n} = (\sin \theta \cos \varphi, \sin \theta \sin \varphi, \cos \theta)$ and the rotation angle $\alpha$.
The initial representation is defined by applying the same rotation to each atom,
\begin{align*}
    \rho(U) = \left( \cos \frac{\alpha}{2} \, \mathbb{I} - i \sin \frac{\alpha}{2} \, \vec{\hat{n}} \cdot \boldsymbol{\sigma} \right)^{\otimes n},
\end{align*}
where the Pauli matrices $\sigma^x$, $\sigma^y$, $\sigma^z$ act on the two-level system $\{ \ket{0}, \ket{1} \}$.
Integration over the Haar measure can be written as (see section 4.1 of reference~\cite{sternberg1994group})
\begin{equation}
    \int \di U = \frac{1}{(2\pi)^2} \int_0^{2\pi} \di \alpha \sin^2 \frac{\alpha}{2} \int_{-1}^1 \di(\cos \theta) \int_0^{2\pi} \di\varphi .
\end{equation}
The dimension of the irreducible representation with total angular momentum $t$ is $d_t = 2 t + 1$ and the character is given by (see example 12.23 of reference~\cite{hall2015lie})
\begin{align*}
    \chi_t(U) &= \frac{\sin \left[ (2t+1)\alpha/2 \right]}{\sin(\alpha/2)} .
\end{align*}

\begin{table}
    \caption{Pulse parameters to prepare a GHZ state of $n = 2$, $3$, and $4$ atoms. The detuning is zero for each pulse and the phase of drive is constant. The pulses are ordered to prepare the GHZ state from the $\ket{0}^{\otimes n}$ atoms, which is opposite to the order in which the parameters were computed.}
    \begin{tabular}{|c c c |} 
     \hline
     \# atoms & pulse \# & $|\Omega | t/\pi $   \\ [0.5ex] 
     \hline
     2 & 1 & $0.90635$  \\ [1ex]
     & 2 & $1.42788$ \\ [1ex]
     & 3 & 1 \\ [1ex]
     & 4 & 0.70711  \\ [1ex]
     \hline
     3 & 1 & $0.71439$  \\ [1ex]
     & 2 & $1.40051$  \\ [1ex]
     & 3 & $0.86508$ \\ [1ex]
     & 4 & 0.70711  \\ [1ex]
     & 5 & 1  \\ [1ex]
     & 6 & 0.57735  \\ [1ex] 
     \hline
    4 & 1 & $0.56629$   \\ [1ex]
    & 2 & $1.30474$  \\ [1ex]
    & 3 & $0.80500$ \\ [1ex]
    & 4 & $0.70711$  \\ [1ex]
    & 5 & 0.70711 \\ [1ex]
    & 6 & 0.57735 \\ [1ex]
    & 7 & 1  \\ [1ex]
    & 8 & 0.5 \\ [1ex] 
    \hline
    \end{tabular}
    \label{tab:table}
\end{table}

We next apply these expressions to a product state with $n_0$ atoms in the state $\ket{0}$ and the remaining $n_1$ atoms in the $\ket{1}$ state. This yields
\begin{align}
    \langle \psi | P_t | \psi \rangle &= \frac{2t+1}{2\pi} \int_0^{2\pi} \di \alpha \, \sin \frac{\alpha}{2} \sin \left[ (2t+1) \alpha / 2 \right] \nonumber\\
    &\times \int_{-1}^1 \di (\cos \theta) \left( \cos \frac{\alpha}{2} - i \sin \frac{\alpha}{2} \cos \theta \right)^{n_0} \nonumber\\
    & \hspace{1.5cm} \times \left( \cos \frac{\alpha}{2} + i \sin \frac{\alpha}{2} \cos \theta \right)^{n_1} .
\end{align}
The integral over $\cos \theta$ can be carried out by parts, resulting in a sum of elementary integrals over $\alpha$. Assuming that $n_0 \geq n_1$, we arrive at
\begin{align}
     \langle \psi | P_t | \psi \rangle &= (2t+1)\sum_{k=0}^{n_1} \frac{n_0!n_1!}{k!(n+1-k)!} \delta_{k,n/2-t} .
\end{align}
The sum is zero unless $ 0 \leq \frac{n}{2}-t \leq n_1$ or, equivalently, $\frac{1}{2}(n_0 - n_1) \leq t \leq \frac{n}{2}$. By noting that the answer must be symmetric under the exchange of $n_0$ and $n_1$, we finally obtain
\begin{align}
    \langle \psi | P_t | \psi \rangle = (2t+1)\frac{n_0!n_1!}{(n/2-t)!(n/2+t+1)!}
\end{align}
if $\frac{1}{2} |n_0-n_1| \leq t \leq \frac{n}{2}$ and $\braket{\psi | P_t | \psi} = 0$ otherwise. \Eqref{eq:weight} follows by substituting in $t = p/2$ and noting that $n = p + 2 q$. The latter equation holds because $\ket{\psi}$ contains no Rydberg excitation by assumption, which implies that only irreducible representations with $\lambda_3 = 0$ contribute.

\section{State preparation\label{app:preparation}}

Following the strategy outlined in section \ref{sec:preparation}, one can prepare any permutation-invariant state from the initial state $\ket{0}^{\otimes n}$ by applying a sequence of alternating pulses that drive the transitions $|0\rangle \leftrightarrow |r\rangle$ and $|1\rangle \leftrightarrow |r\rangle$. To determine the pulse parameters, we start from the target state and choose each pulse such that it empties the leftmost occupied node. After at most $2 n$ pulses, the entire population will thus be moved to $\ket{0}^{\otimes n}$. To prepare the target state, we simply time-reverse the pulses.

\begin{figure}
    \centering
    \includegraphics{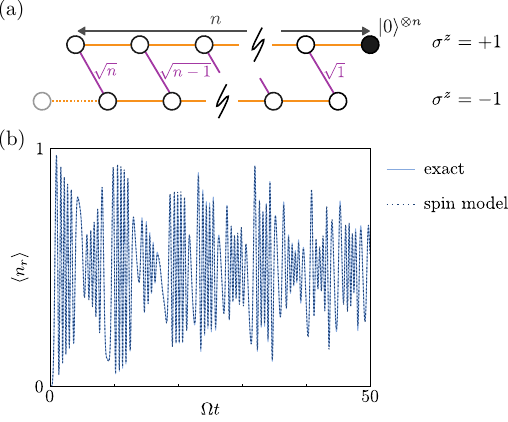}
    \caption{(a)~To construct a spin model for the initial state $\ket{0}^{\otimes n}$, we add an additional, unphysical state to the left of the weight diagram (gray circle connected to dashed line). (b)~Exact evolution and dynamics under the spin model, \eqref{eq:spin_model2}, of $n = 300$ atoms with $\Omega_1 = \Omega_2 = \Omega$, $\Delta_1 = \Omega$, and $\Delta_2 = 0$.}
    \label{fig:figs2}
\end{figure}

Determining the parameters of each pulse is computationally straightforward since each pulse only couples pairs of basis states (see \figref{fig:stateprep}). We focus on the two-level system that includes the leftmost occupied node. Given the amplitudes of these two basis states and the corresponding (unnormalized) Bloch vector, we can readily find the pulse parameters that rotate the Bloch vector to the desired pole. It is always possible to set the detuning to zero, corresponding to a rotation axis on the equator. The phase of the Rabi frequency must then be chosen such that the axis of rotation is perpendicular to the plane spanned by the initial Bloch vector and the pole. The rotation axis will point along the $y$-axis if the amplitudes are real.

\begin{figure*}
    \centering
    \includegraphics{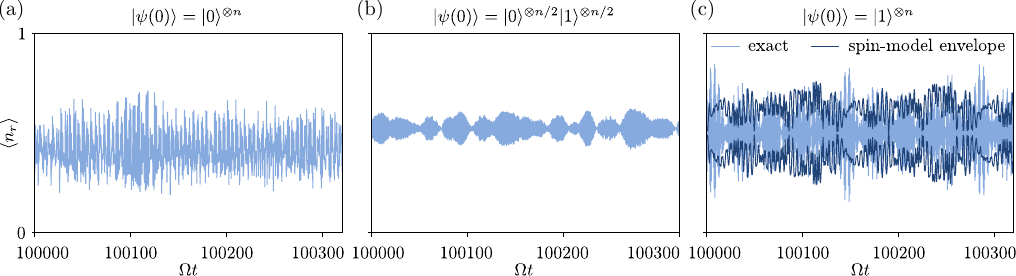}
    \caption{Expected number of Rydberg excitations at late times for the same parameters as in \figref{fig:fig5} ($n = 100$, $\Omega_1 = \Omega_2 = \Omega$, $\Delta_1 = \Omega$, $\Delta_2 = 0$). In panel (c), we also include the envelope computed according to the spin model, \eqref{eq:spin_model}.}
    \label{fig:figs1}
\end{figure*}

We demonstrate this procedure by giving the parameters for the pulses required to create a GHZ state of $2$, $3$, and $4$ atoms in Table~\ref{tab:table}. We note that the rotation angle on the Bloch sphere is given by $|\Omega| t$ times the matrix element of the two-level system of interest. The parameters were obtained for a constant phase of the drive, corresponding to a fixed rotation axis. This is possible because the amplitudes of the state $\left(\ket{0}^{\otimes n} + \ket{1}^{\otimes n}\right)/\sqrt{2}$ are real. We note, however, that this leads to a suboptimal sequence in terms of the total time as certain pulses correspond to a rotation on the Bloch sphere by an angle $\theta > \pi$. These pulses could be shortened by introducing a $\pi$ flip in the phase of the drive, which reduces the rotation angle to $\pi - \theta$.

\section{Equilibration}

\subsection{\texorpdfstring{Spin model for initial state $\ket{0}^{\otimes n}$}{Spin model for initial state |0>}\label{app:model}}

It is also possible to construct a spin model when all atoms are initially in the $\ket{0}$ state. Because the initial state is localized on the right side of the weight diagram, we must add the additional, unphysical state to the left side as shown in \figref{fig:figs2}(a). This ensures that the population in this state remains small for all times at sufficiently large detuning $\Delta_1$.

Making the same approximations as for the $\ket{1}^{\otimes n}$ initial state, the spin model takes the form
\begin{align}
    H_\text{SM}^{(0)} &= \left( \Omega_1 S^x + \Delta_1 S^z \right) \otimes \mathbb{I}_2 + \frac{\Delta_1 + \Delta_2}{2}  \mathbb{I}_{2 s + 1} \otimes \sigma^z \nonumber\\
    &+ \frac{\Omega_2}{2} S^+ M \otimes \sigma^- + \text{h.c.},
    \label{eq:spin_model2}
\end{align}
where $\sigma^- = (\sigma^x - i \sigma^y) / 2$. Here, $M$ is a matrix which is diagonal in the eigenbasis of $S^z$. Denoting the basis states of $S^z$ with eigenvalue $m$ by $\ket{m}$, we have
\begin{equation}
    \braket{m | M | m} = 
    \begin{cases}
        0 & \text{if } m = s\\
        \sqrt{\frac{s - m}{s (s + 1) - m ( m + 1)}} & \text{otherwise}
    \end{cases}.
\end{equation}
The role of the denominator $\sqrt{s(s+1) - m(m+1)}$ is to cancel the matrix element of $S^+$. Hence, the states coupled by the $\ket{1} \leftrightarrow \ket{r}$ drive are connected by a matrix element of magnitude $\sqrt{s - m}$, as indicated by the purple labels in \figref{fig:figs2}(a).

The presence of the operators $S^\pm$ in \eqref{eq:spin_model2} is a key difference compared the spin model for the initial state $\ket{0}^{\otimes n}$. As a consequence, the state does not factorize unlike in \eqref{eq:ansatz}. The two-level system and the large spin become entangled in a more complex fashion and, in particular, do not periodically disentangle. This explains the absence of revivals for the initial state $\ket{0}^{\otimes n}$. Nevertheless, the spin model provides an accurate description of the dynamics as shown in \figref{fig:figs2}(b).

\subsection{Dynamics at very late times\label{app:late}}

In \figref{fig:figs1}, we show the number of Rydberg excitations at late times for the same parameters as in \figref{fig:fig5}. The plots indicate that the revivals do not survive up to infinite times. Nevertheless, the dynamics remain complex and the number of Rydberg oscillations fluctuates significantly even at late times. In \figref{fig:figs1}(c), we also show the envelope of the spin model described in the main text. The agreement with the exact data is poor for this late-time dynamics. This is expected at late times due to the approximations of the spin model.

\end{document}